\documentclass[%
 reprint,
 amsmath,amssymb,
 aps,
pre,
floatfix,
]{revtex4-1}

\usepackage{graphicx}
\usepackage{dcolumn}
\usepackage{bm}
\usepackage{epstopdf}
\usepackage{siunitx}
\usepackage{gensymb}
\usepackage{float} 
\usepackage{amssymb,booktabs,mathtools}

\begin{document}


\title{Graded Index Confined Spin Waves in an Intermediate Domain Wall}

\author{D. Osuna Ruiz *\textsuperscript{1}} 
\email{do278@exeter.ac.uk}
\author{A. P. Hibbins\textsuperscript{1}}
\author{F. Y. Ogrin\textsuperscript{1}}%
 
\affiliation{%
 \textsuperscript{1}Department of Physics and Astronomy, University of Exeter, Exeter EX4 4QL, United Kingdom.
 }%

\date{\today}

\begin{abstract}
We propose a mathematical model for describing propagating confined modes in domain walls of intermediate angle between domains. The proposed model is derived from the linearised Bloch equations of motion and after reasonable assumptions, in the scenario of a thick enough magnetic patch, are accounted. The model shows that there is a clear dependence of the local wavenumber of the confined spin wave on the local angle of the wall and excitation frequency used, which leads to the definition of a local index of refraction \textit{in the wall} as a function of such angle and frequency. Therefore, the model applies to 1-D propagating modes, although it also has physical implications for 2-D scenarios where a domain wall merges with a saturated magnetic region. Micromagnetic simulations are in good agreement with the predictions of the model and also give insight on the effects of curved finite structures may have on the propagating characteristics of spin waves in domain walls.  

\end{abstract}

\pacs{Valid PACS appear here}
\maketitle


\section{\label{sec:level1}introduction}

Due to their low loss and shorter wavelength compared to electromagnetic waves in free space, spin waves are a promising candidate for information carrier in micron and sub-micron scale magnonic circuits \cite{0022-3727-43-26-264001,Karenowska2016,PhysRevApplied.4.047001}. Inhomogeneities such as vortex core have been widely studied as spin waves emitters \cite{article,dieterle19,sluka19}. Once the spin wave is excited, an adequate control on its propagation is key for the developing of circuits that conducts the spin wave through channels. Local excitation of spin waves and its spatial confinement has been widely studied in terms of local ferromagnetic resonances due to inhomogeneities \cite{davies15,davies17,mushenok17}, confinement along the edges \cite{aliev11,Lara2017InformationPI}, along domain walls and by domain wall natural fluctuations modes or so called Winter magnons \cite{aliev11b}. Domain walls do actually act as natural channels for spin waves due to the energy well. More importantly, Winter magnons are very useful for efficiently channeling in a wide range of frequencies since they are gapless modes \cite{winter}. 
\par Graded index media for wave propagation have been widely studied specially in electromagnetics, in the field known as transformation optics \cite{Kang08,rinkevich15,pen00}. The development of structures ranging from nanometre size to centimetres, have proved interesting properties not shown in nature, leading to the application of novel graded-index (GRIN) lenses in an extremely wide range of frequencies, from microwaves to visible light. 
\par In the field of magnonics, graded index magnetic media serve to similar purposes, taking advantage of the high anisotropic behaviour of spin waves \cite{davies15,DaviesKruglyak15}. For example, tailoring the spin wave propagation in magnetic domains, allow the development of lenses for spin waves\cite{Toedt2016DesignAC,white19,white18}. Regarding uni-directional propagation in domain walls, previous studies have dealt mainly with redirection and steering the spin wave path \cite{hamalainen18,alb18}, inducing phase changes \cite{bayer05,wangwang2015domain} or non-reciprocal paths by means of non-linear effects \cite{garcia15}.
\par However, anisotropic magnetic media can not only modify direction, intensity, or temporal frequency of the spin wave. Spatial frequency modulation or ‘spatial chirping’ is a technique widely used in telecommunication engineering and photonics, for example, to use in fibre-Bragg gratings or other chirped mirrors as filters, where wavenumber or equivalently, wavelength ($\lambda=2\pi/k$), spatially changes. Analysis of spatially chirped signals can be extended even to the processing of images where periodic features are visualised in perspective. The graded-index technique for EM waves could also have its equivalent for spin waves, and therefore, mathematical tools are required to control this feature. Using the non-uniform demagnetising field in a saturated YIG non-ellipsoidal rod, the pioneer work from Schlömann \cite{schlomann64} for Backward Volume Spin waves and from Stancil \cite{stancilmorg83} for Surface Spin waves and further experimental results using spatially varying external fields \cite{smith08} confirmed the realization of this technique for spin waves. A change of wavelength has also been observed by using tapered saturated magnonic waveguides \cite{demidov11}. In this article, we demonstrate a spatially dependent wavenumber (spatial dispersion) of confined modes in domain walls (and therefore, in non-saturated films), providing with an equation that allows to model their local wavenumber and propagation properties. Micromagnetic simulations on a specifically designed shape of patch, that retains in remanence a single intermediate domain wall of variable angle between two vortex cores, show good agreement with our proposed model.

\section{\label{sec:level2}Materials and methods}

To obtain insight into the dynamics, we performed a set of micromagnetic simulations using Mumax3 \cite{doi:10.1063/1.4899186}. We simulated a \lq rounded bow-tie' shaped patch (see Fig. 3(a)) of 6000 nm length, 80 nm thickness (${t}$) and 2000 nm diameter ($d$) of the circumscribed circles at the ends, with the typical material parameters of permalloy at room temperature with saturation magnetization $M_s$ $= 7.2$ $\times$ $10^5$ Am$^{-1}$, exchange constant $A_{ex} = $ $1.3$ $\times$  $10^{-11}$ Jm$^{-1}$ and Gilbert damping constant $\alpha$ $=$ $0.008$ for a weighted average of iron and nickel. The grid was discretized in the ${x, y, z}$ space into 1536 $\times$ 512 $\times$ 16 cells. The cell size along ${x}$ and ${y}$ was 3.9 nm, while the cell size along ${z}$ was fixed to 4 nm. The cell size along three dimensions is always kept smaller than the exchange length of permalloy (5.3 nm). The number of cells was chosen to be powers of 2 for sake of computational efficiency. We also set a \lq smooth edges' condition with value 8.
A key point in micromagnetic simulations is to achieve a stable equilibrium magnetization state. We first set a double vortex state with polarity and \lq vorticity' numbers of  (1, $-$1) and (1,$+$1) and then executed the simulation with a high damping ($\alpha$ $=$ 1) ) to relax the magnetization until the maximum  torque (\lq maxtorque' parameter  in Mumax3), which describes the maximum torque/$\gamma$ over all cells, where $\gamma$ is the gyromagnetic ratio of the material), reached {$10^{-7}$} T indicating convergence and the achievement of a magnetization equilibrium state. The typical time to achieve the equilibrium state was 100 ns. Note that this value has no direct physical meaning due to the artificial high damping. Once the ground state was obtained, damping was set back to original ($\alpha$ $=$ 0.008),  relaxation process was repeated, the spin configuration was recorded as the ground state of the sample and then used for the simulations with the dynamic activation. For analyzing time evolution of the magnetic signal, we apply a continuous wave excitation with a magnetic field $B$ at a specific frequency {$f_{0}$} in the first vortex core region only,

\begin{gather}
B(t) = B_{0}\text{sin}(2\pi f_{0} t)), 
\end{gather}

and each mode is excited with a relatively small oscillating field, $B_{0}=0.3$ mT.
A sampling period of {$T_{s} = $25 ps} was used, recording up to 300 simulated snap-shots in space and time, only after the steady state is reached. With these parameters, the time window of observation of the spin waves propagating in the domain wall covers up to 7.5 ns.

\par To numerically validate the change in wavelength predicted by the model, we run micromagnetic simulations with an excitation frequency of 1.5 GHz and 3 GHz so Winter’s magnons \cite{winter} can be efficiently launched from the core region and travel along the domain wall \cite{garcia15}. Numerical results for the spatially dependent wavelength ($\lambda$(x)) are obtained from the time-averaged spectrograms of the channelled spin wave profiles using a Hanning window of 128 FFT points and 100 overlapping points. Since the spatial wavelength is constantly changing along the domain wall length, a fixed width of the window will introduce a trade-off between spatial frequency resolution and position accuracy in the x-direction. A wider spatial window yields less position accuracy to the wavenumber and a very narrow window cannot properly resolve the spatial frequency at large wavelengths, leading to spatial frequency leakage per frequency bin. The chosen parameters of window width and overlapping points are obtained after an optimisation process considering several different outcomes. Another solution could be using a width-variable window as described in \cite{nasir16}.

\section{Description of the model}

As a first step to a reliable model and similarly to what is done in Ref.\cite{bayer05}, we search for an expression of a spatially dependent wavelength for a spin wave travelling in a domain wall, based on a Wentzel-Kramers-Brillouin (WKB) approximation \cite{Zil95}. Assuming a Neel-type domain wall along the x-direction, this implies that dynamic magnetization components at its centre can be expressed as: $m_{\text{x}}=m_0e^{i\omega t}, m_{\text{y}}=M_{\text{s}}, m_{\text{z}}=m_0e^{i\omega t}e^{ik(x) x}$. As it will be shown later in this section, this description of static and dynamic magnetization is still approximately correct even for a Bloch wall of an arbitrary angle and suitable, at least, for the in-plane component of magnetization. If we assume an internal magnetic field perpendicular to the wall and only related to dipolar and exchange interactions, $H_{\text{i}}=(H_{\text{d}}+H_{\text{ex}})\boldsymbol{y}$, where the demagnetising field is $H_{\text{d}}=H_{\text{d}}(x)$ and the exchange field is $H_{\text{ex}}=2A_{\text{ex}}\nabla^2m(x)/(\mu_{0}M_{\text{s}})$. As in Ref. \cite{bayer05}, in order to find an expression for a spatial dependent wavenumber $k(x)$ accounting for the demagnetising field magnitude $H_{\text{d}}(x)=|H_{\text{d}}(x)|$, we define a compact expression for magnetisation $\phi=m_{\text{x}}+im_{\text{z}}$, assuming $\phi \sim \phi_{0}e^{ik(x) x}$ and neglecting the imaginary terms in order to obtain real-valued solutions, we reduce the linearised Bloch equations of motion to the following first-order, non-linear differential equation, 

\begin{gather}
\frac{\omega_0}{|\gamma|} = |H_{\text{d}}(x)| + \frac{2A_{\text{ex}}}{\mu_0 M_\text{s}}\left( x^2\left( \frac{dk}{dx}\right)^2+2x\frac{dk}{dx}+k^2\right), 
\end{gather}
where $\omega_0$ is the excitation frequency and $\gamma$ the gyromagnetic ratio. As a first approach to solve this differential equation, we can naively assume that variations of spin wave wavelength along the domain wall will be smooth, since no sudden changes in the demagnetizing field along the longitudinal direction are expected in a straight domain wall configuration. This implies that the first derivatives with respect to x can be neglected. Similarly to Schlömann’s work \cite{schlomann64}, clearing $k(x)$ in Eq.(2) gives an approximate description for a spatially dependent wavenumber, as a function of the module of the demagnetizing field in the wall,
\begin{gather}
k(x) = \sqrt{k_{0}^2- \frac{\mu_0 M_\text{s}}{2A_{\text{ex}}}|H_{\text{d}}(x)|}, 
\end{gather}
where $k_0=\sqrt{\frac{\mu_{0}M_{\text{s}}}{2A_{\text{ex}}\gamma}\omega_{0}}$ is the wavenumber for a confined spin wave of frequency $\omega_0$ when the transversal in-plane demagnetizing field is zero, or in other words, in a 180 degrees Bloch domain wall \cite{garcia15}. 
\par Finding a general demagnetizing field expression in a non-saturated ferromagnet of a non-ellipsoidal, arbitrary shape can be a very tough task. To address this, we first provide a magnetostatic first approach for such a scenario based on how in-plane magnetization transversally changes through the wall, this is, from one domain into the other. In other words, we aim to establish a link between the Eq. (3) and the angle ($\alpha$) between domains.

\begin{figure}[ht]
\centering 
\includegraphics[trim=0cm 0cm 0cm 0cm, clip=true, width=8cm]{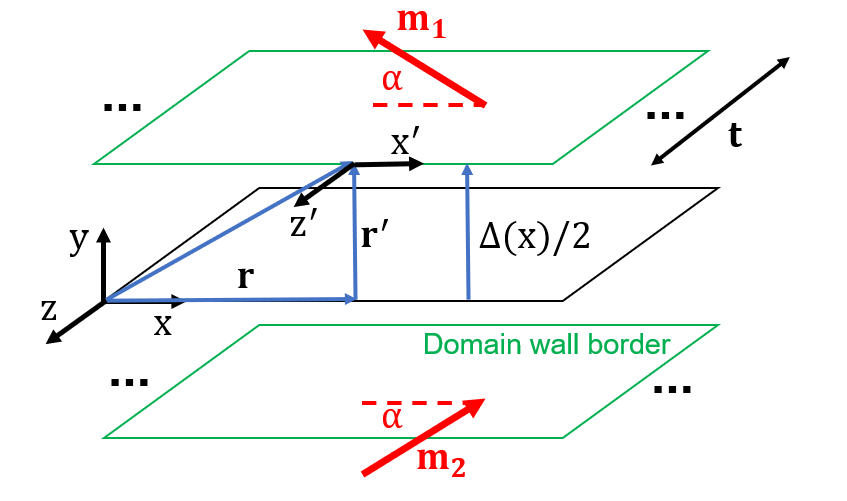}
\caption{Schematic of a domain wall in terms of the domain magnetisation (red arrows) and their angle ($\alpha$) with respect to the domain wall of initial width $\Delta(0)=\Delta_0$ in a sample of thickness $t$. The domain wall borders (green areas) and the domain wall centre (black area) are shown. The absolute and relative coordinate systems, chosen for the calculations, are also shown.}  \label{Fig2}
\end{figure}
\par Let us assume the scenario from Fig. 1 for the domain wall of initially constant width $\Delta(x)=\Delta_0$ and angle $\alpha$ that magnetisation \textbf{$m_{1,2}$} makes with the straight domain wall. Solving the static Landau-Lifshitz-Gilbert equation yields the Walker’s profiles across the domain wall \cite{stancilpra}. Below a critical angle $\alpha_{\text{c}}$, a mixed Bloch-Neel behaviour of the wall is obtained, where the Neel component is always dominant if $t\sim\Delta_0$ \cite{torok65}. A sin($\alpha$)-dependent coefficient is naively added to the magnetisation in-plane components transverse to the wall ($m_{\text{y}}$) to consistently model the angle dependence of a mixed wall. Following similar calculations to those shown in Ref. \cite{hubert98}, chapter 3.9 (a more detailed mathematical description of the derivation of the following equations can be found in Supplemental Material (1)), and considering that $M_{s}=|\textbf{m}|$  must be satisfied for every y-position across the domain wall, the three components of magnetisation in the wall region can be regarded as,

\begin{gather}
m_{\text{x}}=M_{\text{s}}\sqrt{1-\text{sin}^2(\alpha)\text{sech}^2\left( \frac{\text{y}}{\Delta_0}\right)-\frac{\text{cos}^2(\alpha)\xi(\alpha,\text{y})^2}{(1-\text{sin}(\alpha))^2}}
\\
m_{\text{y}}=M_{\text{s}}\text{sin}(\alpha)\text{sech}\left(\frac{\text{y}}{\Delta_{0}}\right)
\\
m_{\text{z}}= M_{\text{s}}\frac{\text{cos}(\alpha)}{1-\text{sin}(\alpha)}\xi(\alpha,\text{y})
\end{gather}
where
\begin{gather}
\xi(\alpha,\text{y})=\frac{1+\text{sin}(\alpha)\text{cosh}\left(\text{cos}(\alpha)\frac{\text{y}}{\Delta_0}\right)}{\text{sin}(\alpha)+\text{cosh}\left(\text{cos}(\alpha)\frac{\text{y}}{\Delta_0}\right)} -\text{sin}(\alpha) 
\end{gather}
Following a magnetostatic approach and assuming that $\frac{dm_{\text{x}}}{dx}\approx 0$  and $\frac{dm_{\text{z}}}{dz}\approx 0$ , a bulk magnetic density charge ($\rho_\text{m}(\textbf{r}')=\rho_\text{m}(\text{y})=\nabla \textbf{M}(\text{y})=\frac{dm_{\text{y}}}{dy}$) can be found,

\begin{gather}
\rho_{\text{m}}(\textbf{r}')=\frac{-M_{\text{s}}\text{sin}(\alpha)\text{tanh}\left( \frac{\text{y}}{\Delta_0}\right)\text{sech}\left( \frac{\text{y}}{\Delta_0}\right)}{\Delta_0}
\end{gather}
The antiderivative of the latter expression effectively retrieves $m_\text{y}$ (Eq. (5)), which implies a \lq slow' variation of the demagnetising field at the central region of the domain wall ($\text{y} \rightarrow 0$), in accordance with the y-dependence of the \lq sech' function. Notice that, if $\alpha = \alpha(x)$, $\Delta_0 = \Delta_0(x)$ and a cross-tie wall were formed, the term $\frac{dm_{\text{x}}}{dx}$ is not necessarily zero, and therefore, it should count into the expression of the bulk magnetic density charge. However, for the sake of simplicity, in this case we consider slow variations of the domain angle and domain wall’s width along the domain wall’s length ($\frac{d\alpha}{dx}\approx 0$  and $\frac{d\Delta_0}{dx}\approx 0$) and no formation of a cross-tie profile. This implies that $\frac{dm_{\text{x}}}{dx}\approx 0$. Also, a thick enough patch is considered, so at the central region of the patch, far enough from the surface pinning effects, variations of magnetization across the thickness can also be neglected ($\frac{dm_{\text{z}}}{dz}\approx 0$) even considering the formation of a cross-tie wall \cite{metlov01}. Under the above mentioned assumptions and due to the conservation of the transverse component in the domain wall border (following on the schematic from Fig. 1), surface magnetic charges can also be considered negligible ($\sigma_\text{m}=\text{m}_{\text{y},1}-\text{m}_{\text{y},2}=0$). Choosing the right integration volume under reasonable assumptions, for example a rectangular prism of chosen dimensions equivalent to the domain wall width ($\Delta_0$) in the x- and y-directions and to the thickness ($t$) in the z-direction, an expression of the demagnetising field can be found. This expression, derived from the magnetostatic potential obtained in turn from the defined magnetic density charge, Eq. (8) (see Supplemental Material (2)), is,

\begin{gather}
H_{\text{d}}(\text{y})=\frac{-tM_{\text{s}}\text{sin}(\alpha)\text{tanh}\left( \frac{\text{y}}{\Delta_0}\right)\text{sech}\left( \frac{\text{y}}{\Delta_0}\right)}{4\pi \text{y}}
\end{gather}
where $t$ is the thickness of the magnetic patch and $\Delta_0$ the \lq constant' domain wall width. A more detailed derivation of Eq.(9) is shown in the Supplemental Material (2). Note that Eq. (9) is only valid for values of $y$ in the wall region ($\frac{-\Delta_0}{2}<y<\frac{\Delta_0}{2}$). Also, Eq. (9) yields a numerical indeterminate at the centre of the wall although this is not a physically realisable solution. At the midwidth (taking the limit $\text{y} \rightarrow 0$) of the domain wall, where the confinement of the mode is strongest (in accordance with the maximum of the \lq sech' function), and assuming $t\sim\Delta_0$ in a thick enough sample, Eq. (9) finally leads to 

\begin{gather}
H_{\text{d}}(\text{y})=\frac{-M_{\text{s}}\text{sin}(\alpha)}{4\pi}
\end{gather}
This equation allows to express the demagnetizing field perpendicular to the wall, in terms of the arbitrary angle α of magnetization between the magnetic domains and the wall. Moreover, it is also consistent with the assumption of a dominant Neel component in the domain wall profile.  

\par After obtaining the compact expression from Eq.(10), it is worth noting that at an angle of $\alpha=\pi/2$ there is, by definition, no domain wall since magnetisation is \lq continuous'. The wall width is by definition \lq zero' and the model no longer applies. The competition between dipolar and exchange energies means that this angle α will also determine the width of the domain wall. In other words, finding an expression for the demagnetising field as a function of the domain angles and the domain wall width is a self-consistent problem. In order to find. In order to find a more complete and exact expression for $H_{\text{d}}(x)$, an additional demagnetising field should be derived from the respective magnetic potential, again derived from the magnetic density charges defined by $\frac{dm_{\text{x}}}{dx}$. These considerations along with a variable domain wall width $\Delta(x)$ add more complexity to the model. However, as described before, the following reasonable assumptions can be made for sake of simplicity in our approximate model: (1) Since the maximum intensity of the confined mode will be at the centre of the domain wall ($\text{y} \rightarrow 0$), the quotient $\frac{\text{y}}{\Delta_0}$ is very small regardless of the domain wall width while $\Delta_0>0$, or equivalently, for $\alpha<\pi/2$. Under this condition is where the proposed model is intended to be used. Due to the equivalence between $\frac{\text{y}}{\Delta_0}$ and \lq y' when  $\text{y} \rightarrow 0$, replacing the first term by just \lq y' in the equation avoids the dependence on the wall width ($\frac{\text{y}}{\Delta_0} := \text{y}$)); and (2), the main assumption for these results is that the demagnetising field is orientated, almost fully in-plane and perpendicular to the domain wall. This assumption is not far from reality, since it has been proved that, in intermediate domain walls (between 180°-Bloch and Neel wall), the Neel component is dominant. This behaviour is observed in arbitrary α-Bloch walls where $\alpha<180$°. In fact, above a critical angle, the out-of-plane component of magnetisation vanishes, and the wall becomes essentially a Neel wall \cite{olson67}. Not only that, argument is even more convincing for thicker samples, since the critical angle reduces when increasing thickness almost reaching zero when $t\sim\Delta_0$ \cite{torok65}. Eq.(10) is consistent with these results, since it also reduces in magnitude as $\alpha$ reduces.
\par After all these considerations, we need to stress, once again, that finding an exact solution for a local demagnetising field is very often, a very complicated task for non-ellipsoidal shapes, even in saturation or quasi-saturation \cite{osborn45,berkov05}. When not found numerically through micromagnetic simulations, approximate analytical approaches are usually taken under reasonable assumptions \cite{PhysRevB.92.214420}. For all this and realizing that Eq. (10) is a physically consistent model for the defined scenario, we consider it from now on, as a valid first approximation for the transverse demagnetising field along a domain wall of variable domain angle. Therefore, this expression can be combined with Eq. (3) to give a new one where the dependence is now with the variable angle between magnetic domains,

\begin{gather}
k(x) = \sqrt{k_{0}^2- \frac{\mu_0 M_\text{s}^2}{8\pi A_{\text{ex}}}\text{sin}(\alpha(x))}, 
\end{gather}
This equation relies on an initial \textbf{k} ($k_0$ \textbf{x}) which is found to be the wavenumber of a Winter’s magnon for a frequency $\omega_0$ through a 180° domain wall (i.e., when $\alpha=0$). Regarding this, through the dispersion relation of Winter’s magnons, a spatial index of refraction can be defined as $n(x,\omega_0 )=(k(x))⁄k_0$,

\begin{gather}
n(x) =\sqrt{1- \frac{|\gamma|}{\omega_0}|H_{\text{d}}(x)|}= \sqrt{1- \frac{\omega_M}{4\pi \omega_0}\text{sin}(\alpha(x))}, 
\end{gather}
where $\omega_M=\gamma M_{\text{s}}$ and $\omega_0$ is the frequency of a continuous wave excitation. This equation predicts the change in wavenumber (or wavelength) from a given $k_0$ along the domain wall. In other words, a different initial $k_0$  will give different values of local wavenumbers, but always varying in accordance to this model. Therefore, this applies as well to the index of refraction if we assume an index of unity for an arbitrary initial $k_0$. Fig. 2(a) shows a contour plot of the real values of Eq. (12) as a function of the magnitude of $H_{\text{d}}$ and $f=\omega_{0}/2\pi$. It clearly shows, for the right combination for excitation frequency and a generic demagnetising field magnitude, the different values for the real part of the index of refraction (dark blue area shows an entirely imaginary index). Most importantly, it shows how at lower frequencies, the change in wavenumber is more sensitive to the transversal demagnetising field than at higher frequencies. Fig. 2(b) shows the same behaviour even when Eq. (10) is introduced into Eq. (12), which reflects now the dependence on the angle $\alpha$. 

\begin{figure*}[ht]
\center 
\includegraphics[trim=0cm 0cm 0cm 0cm, clip=true, width=\textwidth]{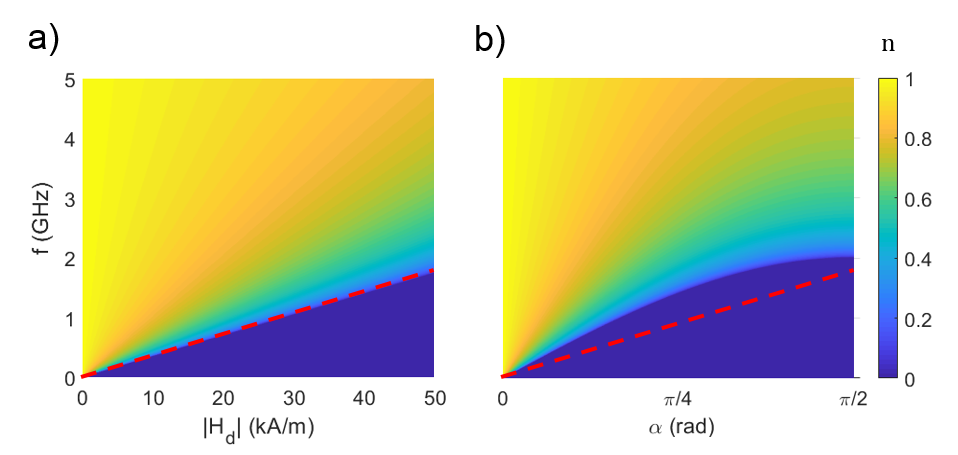}
\caption{Contour plots showing the real part of (a) Eq. (12) as a function of the magnitude of the demagnetising field and (b) Eq. (12) as a function of the angle $\alpha$ for $M_S=720$ kA∙m$^{-1}$, a gyromagnetic ratio $γ=2.2\times 10^5  $Hz(Am)$^{-1}$ and $A_{\text{ex}}=1.3 \times 10^{-11}  $Jm$^{-1}$. Red dashed line with slope $\gamma$ shows the \lq n = 0' condition in (a).}  \label{Fields5}
\end{figure*}

\par It can be inferred from Eq.(12) that, in absence of a biasing field, there is an upper limit for the frequency at which an index of refraction of zero can be obtained. This maximum is reached when the demagnetising field is maximal. In principle, in terms of the magnetic moments’ orientation, this is when $\alpha=\pi/2$. However, notice that, in this situation, the definition of domain wall is meaningless. The scenario would imply the vanishing of the domain wall into a saturated region in the y-direction (i.e., a magnetic domain). In that region, Eq.(12) can be modified by replacing the formerly dominant demagnetising field from the (also former) domain wall by an expression of an effective or internal magnetic field $H_{\text{i}}$. Interestingly, Eq.(12) implies that an index of refraction of zero would be obtained at $\omega_0=|\gamma|H_{\text{i}}$, which is actually the ferromagnetic resonance main mode (i.e., all precessions in phase) of a saturated film for a given internal field $H_{\text{i}}$. This result agrees with the wave perspective of a uniform FMR precession, which lays a wavenumber of $k=0$ (infinite wavelength and $n=0$). Additional simulations (see Supplemental Material (3)), show that at this combination of frequency and effective field, the FMR main mode is not necessarily excited in the saturated region but instead, a spin wave mode with at least $k_{x}=0$ (but not necessarily $k_{y}=0$) can be excited (i.e., a Backward Volume Spin Wave mode). This result still agrees with the perspective of a \lq confined mode' \textit{in the x-direction}, since the model becomes a loose approximation to a scenario where domains or saturated regions are present. In other words, it does not account for modes that may show non-zero wavenumber in a different direction of propagation, such as the y-direction, orthogonal to the original domain wall, although it effectively retrieves the expected zero wavenumber in the x-direction.  More interestingly, the model predicts the existence of, conceptually speaking, \textit{evanescent} spin waves below that \lq pseudo-FMR' frequency condition.

\section{Numerical validation}

\par A mathematical model is only as good as the assumptions. In order to validate our model for the real values obtained from Eq.(12), micromagnetic simulations on the magnetic structure shown in Fig. 3 are performed and recorded for comparison with the analytical model. Since it is inevitable to start from an angle of approximately $\pi/4$ at the source (the vortex core), the shape of this structure has been chosen so because it allows to cover the widest range of values of $\alpha$ and therefore of n, see Fig.4. Therefore, for the shape of this patch, the demagnetizing field transversal to the wall is not constant along its length: The angle $\alpha$ at both sides of the wall changes from $\pi/4$ at the core region to 0 at the centre of the shape and to $\pi/4$ back again. 

\begin{figure}[ht]
\centering 
\includegraphics[trim=0cm 0cm 0cm 0cm, clip=true, width=9cm]{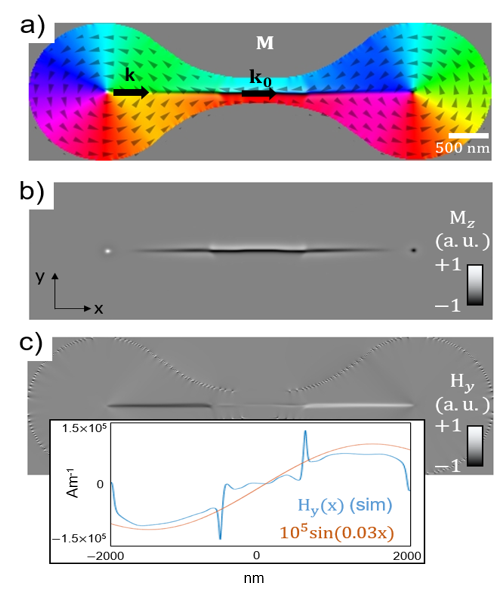}
\caption{(a) Schematic of the proposed structure of 2000 nm × 6000 nm × 80 nm. A Bloch domain wall is induced and left to relax before running dynamic excitations.(b) Normalised out-of-plane component of magnetisation is shown, demonstrating the formation of a Bloch domain wall in the middle of the structure. (c) Normalised in-plane y-component of the demagnetising field, showing a reduction in magnitude in the centre of the structure. Inset in (c) shows the magnitude of the in-plane component of the demagnetising field, perpendicular to the wall, at the center of the wall obtained from micromagnetic simulations (blue curve) and a sinusoidal dependence with x-position between the two vortex core positions (x = 2000 nm and x = $-$2000 nm), in qualitative good agreement with the proposed model from Eq. (6.2.12). The spatial frequency of 0.03 (nm$^{-1}$) is obtained from the spatially-dependent angle α between magnetic moments in the shape (Fig. 6.5(a)).}  \label{Fig3}
\end{figure}
\par Notice that, as reference, an index of refraction of unity midway between the vortex cores, in a 180° Bloch wall, is considered. The reference wavenumber $k_{0}$ is that of a Winter’s magnon along a 180° Bloch wall. For a spin wave propagating from one of the core regions, reduction in $\alpha$ (or equivalently, reduction in the transverse demagnetizing field, see Fig. 3(c)) implies an increase in the local wavenumber. From numerical results on this particular shape, small and smooth variations of the angle are considered far from the core regions, so the assumption $\frac{d\alpha}{dx}\approx 0$  and $\frac{d\Delta_0}{dx}\approx 0$  is still hold and therefore: $\frac{dm}{dy}\approx\frac{dm_{\text{y}}}{dy}$. Close to the core regions, the domain wall angle is large enough to assume the domain wall Neel component to be dominant \cite{olson67} and therefore, $\frac{dm}{dy}\approx\frac{dm_{\text{y}}}{dy}$ is also satisfied at the centre of the domain wall. In other words, the assumptions held in the previous section are still valid as well as the derived model for the demagnetising field transverse to the wall.  

\begin{figure}[ht]
\centering 
\includegraphics[trim=0cm 0cm 0cm 0cm, clip=true, width=8cm]{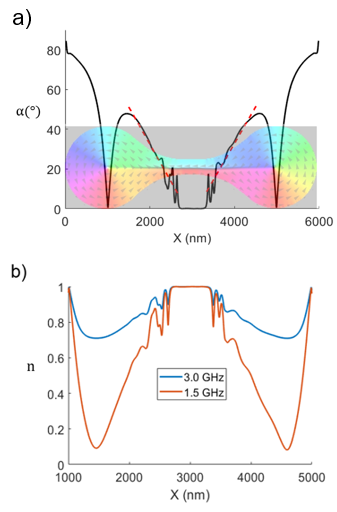}
\caption{(a) Profile of angle $\alpha$ for the structure, calculated as $\text{tan}^{-1}(m_{\text{y}}/m_{\text{x}})$ along a line, parallel to the wall, at 50 nm from the domain wall centre. Red dotted line is a linear fit to the values from the vortex core region to the centre of the structure. (b) Local index of refraction from Eq. (12) with the wavelength of a Winter’s magnon in a 180° Bloch wall as reference for 3 GHz and 1.5 GHz.}  \label{Fig4}
\end{figure}
\par Fig. 4(a) shows the spatial position profile of α parallel to the domain wall at 50 nm from the centre of the wall, calculated as $\text{tan}^{-1}⁡(m_\text{y}/m_\text{x})$, where the components of magnetisation are extracted from numerical simulations and magnetisation is assumed in-plane. The ripples close to the central region come from the numerically obtained equilibrium state of magnetic moments. More specifically, from those in the region where the curved contour is closer to the straight domain wall, which makes magnetization to not smoothly follow the contour neither the domain wall in a straight line but rather, in an apparent ‘zig-zagging’ path as a middle-ground solution. This alters the calculation of the angle $\alpha$ ($\alpha =\text{tan}^{-1}⁡(m_\text{y}/m_\text{x})$), introducing the rippling. Ideally, this should not be happening, and magnetisation should be laying completely in-plane and smoothly following the shape contour even close to the domain wall. Still, the shape of the structure allows to approximate the dependence of angle α with x by a fitting linear dependence $\alpha(x)≈-0.029x+87$, see red dotted line in Fig. 4(a), making it closer to the ideal case. Including this linear fitting into the proposed model for the demagnetising field (Eq. 10) shows also good qualitative agreement with the values obtained from simulations (see inset in Fig. 3(c)). 
\par Quantitatively, result from simulation fits better to a sinusoidal function with a maximum value of $10^5$Am$^{-1}$. While in the same order of magnitude, this value is about three times the maximum of the model from Eq. (10) for $\alpha =\pi/4$, $M_{\text{s}} \text{sin}⁡(\pi/4)/4\pi \approx 0.32 \times 10^5$Am$^{-1}$, (for a saturation magnetisation of $7.2 \times 10^5$Am$^{-1}$). This can be explained by the neglection of other components of the magnetostatic potential in the model (which leads to a \lq weaker' demagnetising field than in reality, due to fewer contributions), that should be accounted for a more accurate description. 
\par Fig. 4(b) shows the analytical index of refraction from Eq. (12) with the obtained values of α as inputs for the excitation frequencies of 1.5 GHz and 3 GHz. This agrees with analytical results from Fig. 2(b), at 3 GHz excitation frequency, which show that the wavevector can be reduced as much as approximately 0.7 times the reference wavevector $k_{0}$ (i.e., $\text{n}\approx0.7$) when the spin wave approaches the core regions (at 900 nm from the vortex core) and $\alpha$ is approximately $\pi/4.8 \approx0.205 \pi$ radians (i.e., 37°, see Fig. 4(a)). 
\par Fig. 5(a) shows the wave profile of the propagating mode and the respective wavelengths found in the respective regions I ($\lambda_{\text{I}}$) and II ($\lambda_{\text{II}}$) (insets). A significant attenuation of the spin wave is observed in region II (the colour intensity in the insets is manually adjusted for ease of comparison), we believe this is originated by the \lq zig-zagging' path of magnetisation in the domains which induces  a pronounced gradient in the local demagnetising field halfway between the two regions (see blue curve in the inset of Fig. 3(c)), modifying in turn the spatially local ferromagnetic resonance \textit{in the wall} and thus presenting low transmission between regions in the wall \cite{wang13,chang18}, as a 1-D analogy of the 2-D scenario explored in Ref.\cite{chang18}. Fig. 5(b) shows the index of refraction profiles for 1.5 GHz and 3 GHz (solid lines) obtained from Eq.(12), with the range of angles found from simulations as input, and numerical results (dots) for the spatial frequency in the x-direction, normalised to the value at 3000 nm (or in other words, the simulated spatial index of refraction). Numerical results confirm what the equation predicted: The change in wavenumber is more pronounced, and more sensitive to the transverse demagnetising field (or equivalently, to the domain angle $\alpha$) when the frequency is smaller. Therefore, the covered graded-index profile is wider at lower frequencies. Also, as expected from the model, the values of the obtained index of refraction from simulations agrees with the predicted values from Eq. (12) with the specified angle $\alpha=37$° at the corresponding position (x = 1900 nm) in the shape ($n\approx0.7$ at 3 GHz and $n\approx0.4$ at 1.5 GHz). An apparent physical anomaly for an excitation frequency of 3 GHz is observed from simulations (blue dots in Fig. 5(b)) halfway between the regions I and II, i.e. around x = 2200 nm. The x-positions near that value actually correspond to the region where the \lq zig-zagging' of magnetisation is more marked, making the simulated value not comparable to the value from the model, that assumes a smooth variation of the demagnetising field. This explains the considerable mismatch at around x = 2200 nm in Fig. 5(b), and shows that the effects of curved edges on propagation in a straight domain wall might be difficult to avoid in certain localised areas in a finite structure and more significant for higher excitation frequencies.
\begin{figure}[ht]
\centering
\includegraphics[trim=0cm 0cm 0cm 0cm, clip=true, width=8cm]{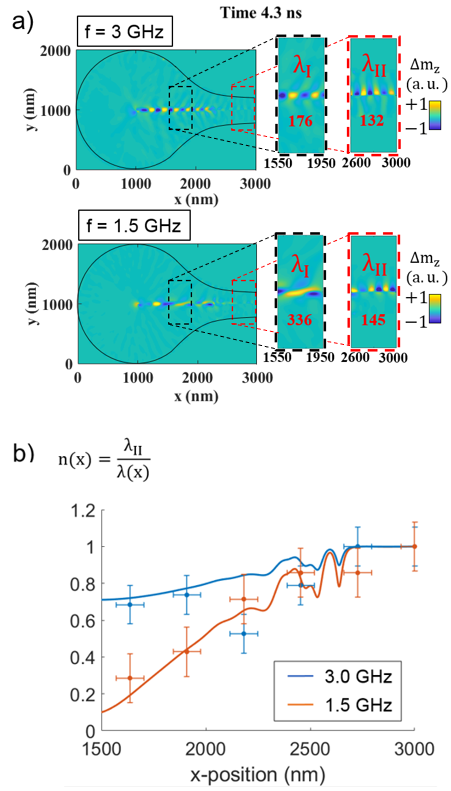}
\caption{(a) Simulated wave profiles for an excitation frequency of 3 GHz and 1.5 GHz, with the chosen regions highlighted in dashed lines (they are not representing the Hanning window): Region I centred around x=1900 nm and region II centred around x=2800 nm, showing the spin wave profile in one half of the structure. Note that, for ease of comparison, the colour scale in the insets for region II is not to scale (it is reduced) to that of region I. The local wavelengths are indicated in the insets, along a length of 400 nm. (b) Results from the analytical model from Eq.(12) for 1.5 GHz and 3 GHz (solid lines) and the ratios $\lambda_{\text{II}}/\lambda(\text{x})$  from micromagnetic simulations (dots) for each frequency and per Hanning window.}  \label{Fig5}
\end{figure}
\par In conclusion, Eq. (11) or its generic form Eq. (3) is proposed as a valid model for finding the local wavenumber (and Eq. (12) for the local index of refraction) of a spin wave along domain walls under the influence of different demagnetising fields or shape contour effects. The model helps to predict how these effects modify the wavelength of a confined mode in a domain wall.

\section{Summary}
\par The main result of this work is the proposed mathematical model for an effective spatial frequency dependence of confined spin waves in an \lq intermediate angle' domain walls as a function of that angle. The equation of the model is derived from the fundamental Bloch equations of motion, an approximated (under reasonable assumptions, similar to those in Ref. \cite{PhysRevB.92.214420} expression of the demagnetising field transversal to the wall, that is applied to a variety of magnonic scenarios. The model can be applied to straight domain walls of variable domain angles, which can be useful as a first approximation to the study of more complex scenarios such as spin waves in confined structures of arbitrary shapes showing magnetic domains. Reciprocally, the connection between the shape of the magnetic \lq patch' and the shape-induced demagnetising field transverse to the wall, allows us to design its shape for a particularly desired channelled spin wave profile. 
\par The model also leads to conditions that have physical meaning such as the FMR main mode frequency or Backward Volume Spin Waves, and therefore further extending the applicability of the model to not localised modes in domain walls.
In summary, an equation for a spatial-dependent wavenumber for spin waves is proposed, performing as a good model for their propagating behaviour in domain walls. This result may help in the development of more complex models for spin wave propagation in non-saturated nanostructures, channelled along domain walls or propagating into magnetic domains.
 
\section{Acknowledgements}
\par This work was supported by EPSRC and the CDT in Metamaterials, University of Exeter. All data created during this research are openly available from the University of Exeter's institutional repository at https://ore.exeter.ac.uk/repository/

\bibliography{library}

\end{document}